\documentclass[aps,twocolumn,showpacs,preprintnumbers]{revtex4}

\usepackage{graphicx}
\usepackage{times}

\begin{document}


\title{Coupled surface polaritons and the Casimir force}

\author{C. Henkel}
\email{Carsten.Henkel@quantum.physik.uni-potsdam.de}
\affiliation{%
Institut f\"ur Physik, Universit\"at Potsdam,
Am Neuen Palais 10, 14469 Potsdam, Germany
}%

\author{K. Joulain}
\altaffiliation[Currently at ]{%
Laboratoire d'\'etudes thermiques,
ENSMA,
86960 Futuroscope Cedex, France.
}
\author{J.-Ph. Mulet}
\altaffiliation[Currently at ]{%
The Institute of Optics, University of Rochester,
Rochester NY 14627, USA.
}
\author{J.-J. Greffet}
\affiliation{%
Laboratoire EM2C, Ecole Centrale Paris,
92295 Ch\^atenay-Malabry CEDEX, France
}%

\date{06 November 2003}

\begin{abstract}
The Casimir force between metallic plates made of realistic materials
is evaluated for distances in the nanometer range.  A spectrum over
real frequencies is introduced and shows narrow peaks due to surface
resonances (plasmon polaritons or phonon polaritons) that are coupled
across the vacuum gap.  We demonstrate that the Casimir force
originates from the attraction (repulsion) due to the corresponding
symmetric (antisymmetric) eigenmodes, respectively.  This picture is
used to derive a simple analytical estimate of the Casimir force at
short distances.  We recover the result known for Drude metals without
absorption and compute the correction for weakly absorbing materials.
\end{abstract}

\maketitle



\section{Introduction}

Van der Waals and Casimir forces are among the few macroscopic
manifestations of vacuum fluctuations.  Since the seminal paper by
Casimir~\cite{Casimir} showing the existence of an attraction between
two perfect conductors separated by a vacuum gap, an abundant
literature has been devoted to this effect.  In particular, the
relevance of retardation, finite conductivity, and finite temperature
have been studied (see, e.g.,~\cite{Schwinger78}).  Exhaustive lists
of references can be found in several review papers such
as~\cite{Plunien,Bordag,Lamoreaux99}.

In the last five years, the interest in Casimir forces has increased
due to the existence of new measurements with improved
accuracy~\cite{Lamoreaux97,Mohideen}. This has challenged theoreticians to
quantify the corrections to the ideal case (zero temperature, perfect
conductors, flat interfaces) that must be taken into account for an
accurate comparison with
experiments~\cite{Klim,Klimchitskaya00,Lambrecht00,Genet00,Tadmor01,Genet03a}.
%
Furthermore, the developments of micro-electromechanical systems
(MEMS), for example, have shown that the Casimir effect is becoming an
issue in nano-engineering~\cite{Chan1,Chan2}. Indeed, these short-range
forces could seriously disturb the performances of MEMS~\cite{Buks}.

From a theoretical point of view, different methods exist to calculate
Casimir forces. Casimir himself~\cite{Casimir} determined the
electromagnetic eigenfrequencies of the system and summed them in
order to obtain the system's zero-point energy. The force is found by
differentiation of this energy with respect to the geometrical
distance separating the bodies~\cite{Casimir,Milloni}. Ingenious
subtraction procedures are often required to obtain a finite value for
the Casimir energy, and realistic dispersive or absorbing materials
can be dealt with using contour integrals over complex
frequencies~\cite{Mostepanenko}.  Another method, used by
Lifshitz~\cite{Lifshitz56}, considers fluctuating currents driven by
thermal or vacuum fluctuations in the whole space. These currents,
whose spatial correlations are known through the fluctuation
dissipation theorem, interact via the electromagnetic fields they
radiate.  The force is
obtained by calculating the flux of the Maxwell stress tensor across a
surface separating the bodies. One thus gets an integral over all
possible partial wave contributions. For two parallel plates 
separated by a vacuum gap, 
for example, the partial waves can be labelled by their
frequency, wave vector parallel to the interface, and polarization.
By using clever contour deformation, Lifshitz greatly
simplified the calculation of the Casimir force integral. The
principal drawback of this approach is that the integrand can no
longer be interpreted as a force spectrum.

In this paper, we use an alternative approach and study the force
integral over real
frequencies and wave vectors. We show for generic materials
(semiconductors and real metals) that in the near-field regime
(separation distance small compared to the wavelengths considered),
the frequency spectrum of the force exhibits peaks located close to
surface-polariton frequencies. These peaks give the essential
contribution to the Casimir force in this regime.  We identify two
types of resonant surface modes, binding and antibinding, that
contribute respectively with an attractive and a repulsive term to the
force. This substantiates early
suggestions~\cite{VanKampen68,Gerlach71} that the Casimir force is due
to surface modes, see also the recent papers by Genet et 
al.~\cite{Genet03a,Genet03b}.

We finally focus on materials whose dielectric function is modeled by
a Lorentzian resonance, including a nonzero absorption.  We are able
to use the qualitative suggestions mentioned above and propose a
quantitative estimation of the Casimir force in terms of coupled
surface resonances. The dominant contribution of these resonances at
nanometer distances allows to perform exactly the integral over the mode
frequencies, whereas the integral over the wave vector is computed to
first order in the absorption.  We show that the respective
contributions of binding/antibinding modes give a simple and
accurate analytical estimate for the short-distance Casimir force,
recovering previous results for nonabsorbing Drude
materials~\cite{Lambrecht00}. In the corresponding Hamaker constant,
we include corrections due to material losses. 
The accuracy of our results is established by comparing to
numerical evaluations of Lifshitz theory, using tabulated
data for the dielectric functions \cite{Palik}.
The paper concludes with a
discussion of possibilities to ``tune'' the Casimir force that are
suggested by our approach.






\section{Surface resonances in the frequency spectrum}

The starting point for our calculation of the Casimir force is Rytov's
theory of fluctuating electrodynamics in absorbing media~\cite{Rytov3}
that has been used by Lifshitz in his seminal
paper~\cite{Lifshitz56}.  This scheme applies to dispersive or
absorbing materials, as long as their dielectric response is
linear. It has also been shown to provide a suitable framework for a
consistent quantization procedure of the macroscopic Maxwell equations
(see~\cite{Scheel00b,Raabe03a} and references therein).

In the following, we focus on the standard geometry of two planar
half-spaces made from identical material (of local complex dielectric
constant~$\varepsilon(\omega)$) and separated by a vacuum gap of width
$d$. In the Rytov-Lifshitz method, the Casimir force is computed from
the expectation value of the Maxwell stress tensor at an arbitary
position in the gap. At zero temperature and after subtraction of
divergent contributions, Lifshitz gets a force per unit area given
by~\cite{Lifshitz56}
\begin{eqnarray}
     && F =
     \int_{0}^{\infty}\!\frac{ {\rm d}\omega }{ 2\pi }
     \int_{0}^{\infty}\!\frac{ {\rm d} u }{ 2\pi } \,
     F(u,\omega)
     \label{eq:1a}
     \\
     && \hspace*{-10mm}
F(u, \omega ) =
	\frac{2\hbar
     \omega^3 u}{c^3}
     \,{\rm Im}\!\left(
     v \!\!\sum_{\mu \,=\,{\rm s},\, {\rm p}}
     \frac{ r_{\mu}^2( u, \omega) \, {\rm e}^{ -2
     \omega  v d } }{
     1 - r_{\mu}^2( u ,\omega ) \, {\rm e}^{ -2
     \omega v d / c } }
     \right)\!
     ,
\label{eq:1}
\end{eqnarray}
where $v = (u^2 - 1)^{1/2}$ (${\rm Im}\,v \le 0$), and $r_{\mu}$ is
the Fresnel reflection coefficient for a plane wave with polarization
$\mu$ and wavevector $K = \omega  u / c$ parallel to the
vacuum-medium interface. We use the convention that an attractive force
corresponds to $F > 0$.
We note that Rytov's approach allows for an easy generalization to
different media held at different nonzero temperatures. The radiation
force on a small polarizable sphere above a heated surface has been
discussed previously in~\cite{Henkel02a}. Results for the
non-equilibrium Casimir force will be reported elsewhere.

Lifshitz evaluated the integrals~(\ref{eq:1a}) by deforming 
integration contour in the complex plane
to arrive at an
integral over imaginary frequencies $\omega = {\rm i}\xi$.  The
integration then requires the continuation of the dielectric function
from real-frequency data to $\varepsilon({\rm i}\xi )$, using
analyticity properties as discussed in~\cite{Klimchitskaya00,Lambrecht00}.  
We follow here a
different route and continue to work with real $\omega$ and $u$,
taking advantage of the fact that Lifshitz' results provides us with 
an expression for the frequency spectrum 
$F( \omega ) = \int F(u,\omega) du /  2\pi$ of the
Casimir force.
Note that the force
spectrum is more difficult to define in a calculation based on mode
summation, see, e.g.,~\cite{Ford1,Ford2}.

\begin{figure}
     \vspace*{10mm}
\begin{center}
\includegraphics[width=\columnwidth,angle=0]{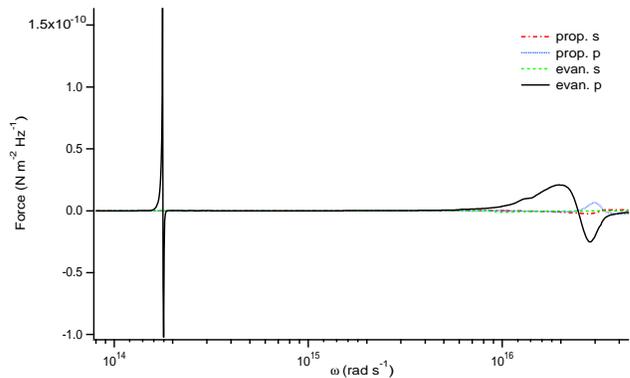}
\end{center}
     \caption[]{Contributions of s and p polarized, propagating and
     evanescent modes to the force spectrum (Eq.\,(\ref{eq:1})
     integrated over the wavevector $u$).  Distance $d =
     10$~nm. Material: SiC, dielectric function taken from
tabulated data~\cite{Palik}.
The corresponding surface resonances
(${\rm Re}\,\varepsilon( \omega ) = -1$) are located at
$1.78\times 10^{14}\,{\rm s}^{-1}$ in the IR and $2.45\times
     10^{16}\,{\rm s}^{-1}$ in the UV.}
\label{fig:1a}
\end{figure}

\begin{figure}
     \vspace*{10mm}
\begin{center}
\includegraphics[width=\columnwidth,angle=0]{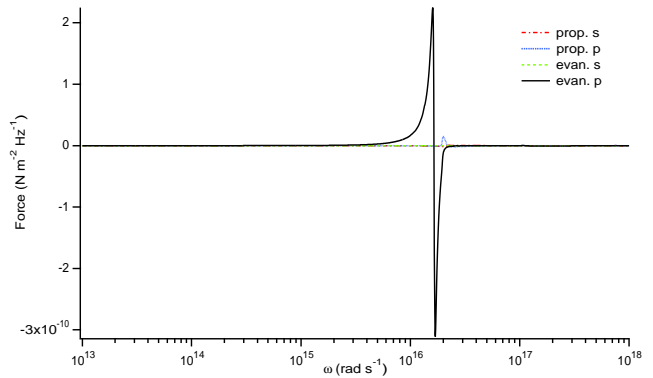}
\end{center}
     \caption[]{%
     Contributions of s and p polarized, propagating and
     evanescent modes to the force spectrum (Eq.\,(\ref{eq:1})
     integrated over the wavevector $u$).  Distance $d =
     10$~nm. Material: aluminum, described by tabulated
optical data~\cite{Palik}.%
}
     \label{fig:1b}
\end{figure}
\begin{figure}
\includegraphics[width=\columnwidth,angle=0]{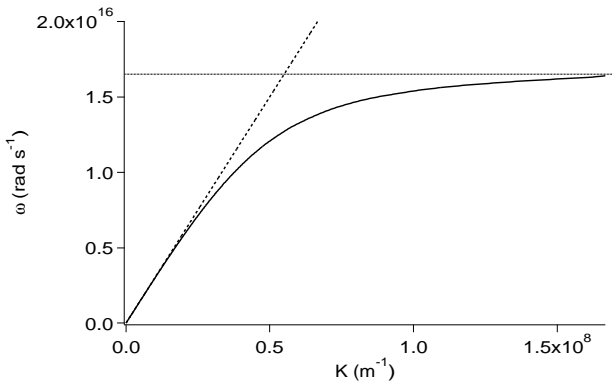}
     \caption{%
Dispersion relation [Eq.\,(\ref{eq:2})] of the surface
     plasmon polariton on a flat interface
     vacuum/aluminum.
     The dielectric function is taken from the data tabulated 
in~\cite{Palik}. 
     We plot the real part of $\omega$ versus
     the real part of the parallel wavevector~$K=u\omega/c$.}
\label{reldisp}
\end{figure}

For a polar material like SiC, the spectrum of the force is dominated
by narrow peaks in the UV and in the IR (Fig.~\ref{fig:1a})
when the distance $d$ is reduced
to the nanometer range.
These peaks can be ascribed to the surface phonon polaritons in the IR
and to surface plasmon polaritons (SPP) in the UV.
The largest contribution comes from the
UV surface plasmon polariton even though larger losses make it 
broader. The large difference between the UV
and the IR contributions in Fig.~\ref{fig:1a} is due to the factor
$\omega^{3}$ in Eq.\,(\ref{eq:1}). In Fig.~\ref{fig:1b}, we plot the
spectrum of the force between two aluminum half-spaces,
using tabulated data for the dielectric function~\cite{Palik}. 
The dominant
contribution to the force is clearly due to the surface plasmon
polaritons. Indeed, the frequency of the peaks corresponds to the
frequency $\Omega$ of the asymptote of the SPP
dispersion relation~\cite{Raether} (see Fig.~\ref{reldisp})
\begin{equation}
u_{\rm SPP} = \sqrt{ \frac{ \varepsilon( \omega )
}{ \varepsilon( \omega ) + 1} },
\label{eq:2}
\end{equation}
where the sign of the square root is chosen such that ${\rm
Re}\,u_{\rm SPP} > 1$. It is seen in Eq.\,(\ref{eq:2}) 
that the
frequency $\Omega$ is given by the condition ${\rm
Re}\,\varepsilon(\Omega)=-1$. This corresponds to a large increase of the
density of states and therefore to a peak in the energy 
density~\cite{Shchegrov,Joulain03}. The polarization dependence of the
force spectrum provides a second argument in favor of a surface
plasmon polariton. 
In Figs.~\ref{fig:1a}, \ref{fig:1b}, we have separated the 
contributions to the spectrum according to the
mode polarization (s or p). The modes in the cavity can further be
classified into evanescent (surface) modes ($u>1$) 
and propagating (guided) modes ($0\leq u\leq 1$).
Among the four contributions
it is seen that the leading one comes from the p-polarized
surface modes, of which the SPP is a special case. 

It is worthwhile pointing
out that for a perfectly conducting metal, the spectrum
of the force would be completely different because of the lack of SPP.
The usual picture of the Casimir effect in that case
is based on the modified density of states for propagating waves
between the two plates. This picture includes only what we have
called guided modes and ignores surface (or evanescent) modes.

We observe from
Figs.\ref{fig:1a},\ref{fig:1b} that the contribution of the force
is either positive or negative depending on the frequency.
We analyze this behaviour in the next section.

\section{Binding and antibinding resonances}

\begin{figure}
\includegraphics*[width=\columnwidth,angle=0]{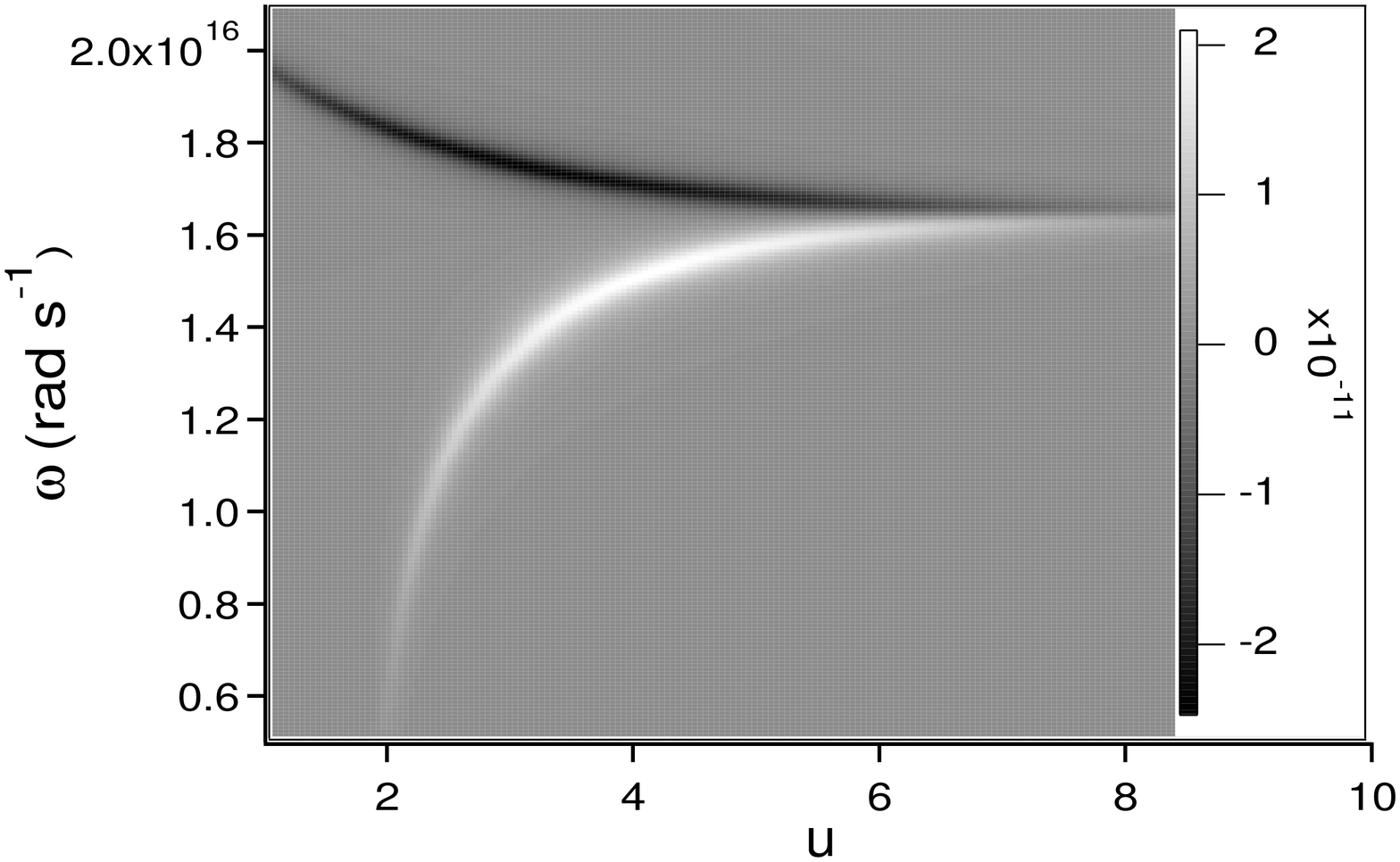}
\includegraphics*[width=\columnwidth,angle=0]{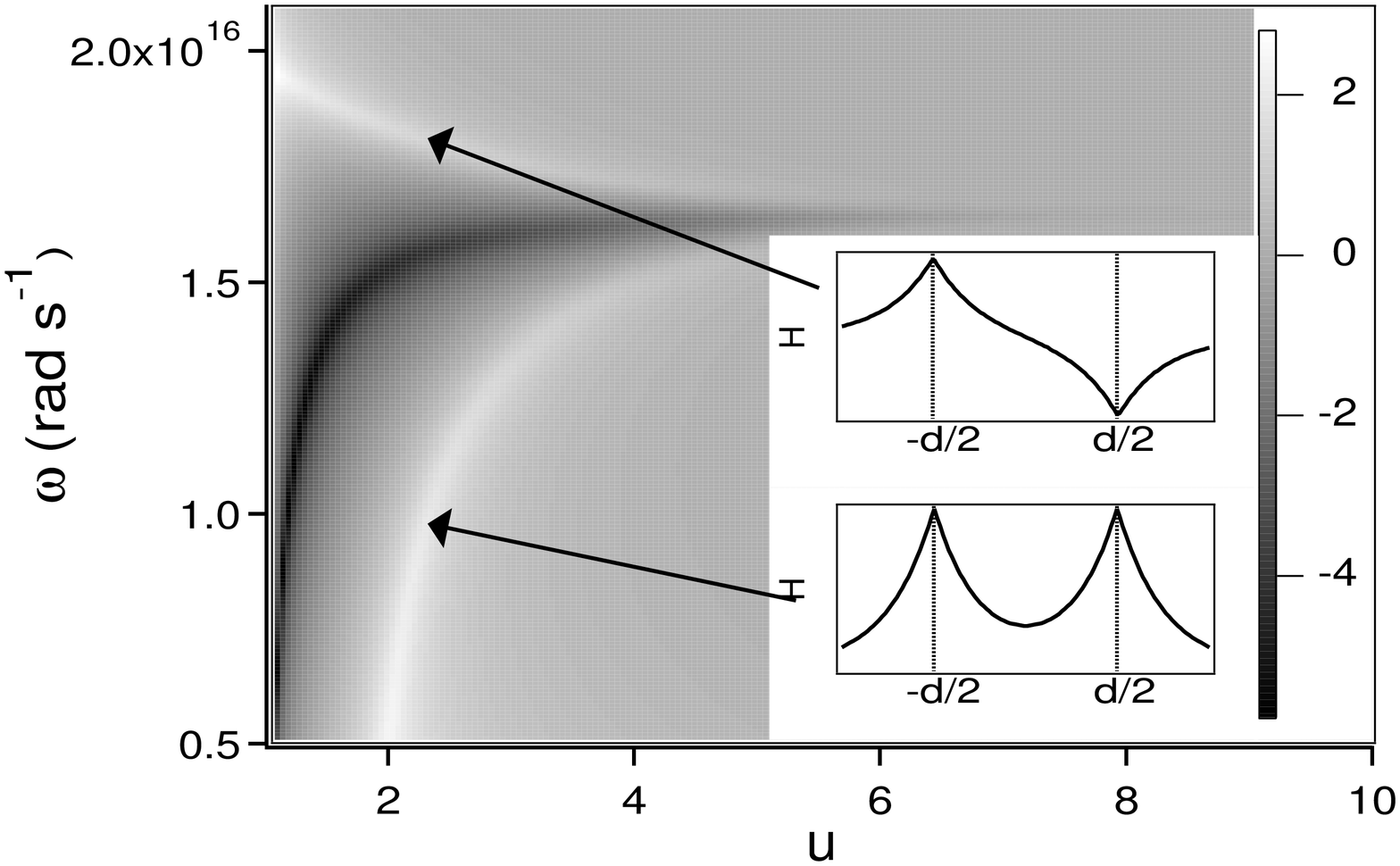}
\caption{%
(a) Wavevector resolved spectrum of the Casimir force 
[Eq.\,(\ref{eq:1})] in the ($u,\omega$) plane 
between two aluminum half
spaces separated by a distance of $10$~nm.  The frequency of the flat
asymptote corresponds to the peaks of the force spectrum
Fig.~\ref{fig:1b}. Light (dark) areas: attractive (repulsive) force.
 (b) Resonant denominator $1/|1 - r_{\rm p}^2 
\,{\rm e}^{ - 2 \omega v d / c }|^2$ in the ($u,\omega$) plane,
the grayscale giving the logarithm to base 10. The dispersion relation
of the coupled surface resonance corresponds to the light areas;
dark area: dispersion relation for a single interface [Eq.\,(\ref{eq:2})].
Dielectric function extracted from tabulated data~\cite{Palik}.
The inset sketches the magnetic field of the coupled surface resonances
(antisymmetric and symmetric combinations).%
}
\label{fig:2}
\end{figure}

In order to further analyze the role of SPPs for the Casimir force, we
plot in Fig.~\ref{fig:2}a the integrand $F(u,\omega)$ as given by
Eq.\,(\ref{eq:1}) for two aluminum half-spaces separated by a distance
of $d = 10$~nm. Two branches emerge with dominant contributions, the
higher-frequency branch yielding a negative contribution whereas the
lower branch gives a positive (attractive) contribution.
These two branches are reminiscent of the dispersion
relation of a SPP on a two interfaces system.  It is given by the
complex poles of the reflection factor of the two interfaces system in
the ($u,\omega$) plane:
\begin{equation}
1-r_{\rm p}^{2}\,{\rm e}^{-2 \omega  v d / c}=0
\label{eq:spp-pole}
\end{equation}
In order to illustrate the influence of the SPP dispersion
relation
on the force, we plot in Fig.~\ref{fig:2}b 
the quantity $1/|1-r_{\rm p}^{2}\,{\rm e}^{-2
\omega  v d / c}|^{2}$  in the real
$(u,\omega)$ plane. Comparing Figs.~\ref{fig:2}b and~\ref{fig:2}a, 
it is clearly seen that the main contribution to
the force is due to the SPP.  In addition, we observe on
Fig.~\ref{fig:2}b a dark line which corresponds to minima of
$1/|1-r_{\rm p}^{2}\,{\rm e}^{-2 \omega  v d / c}|^{2}$.  This can
be attributed to very large values of the 
reflection factor $r_{\rm p}$.  Thus, the dark line is
the dispersion relation of the SPP on a single flat interface.
Note that the Casimir force shows no prominent feature in
this region. 

\begin{figure}
\includegraphics*[width=\columnwidth,angle=0]{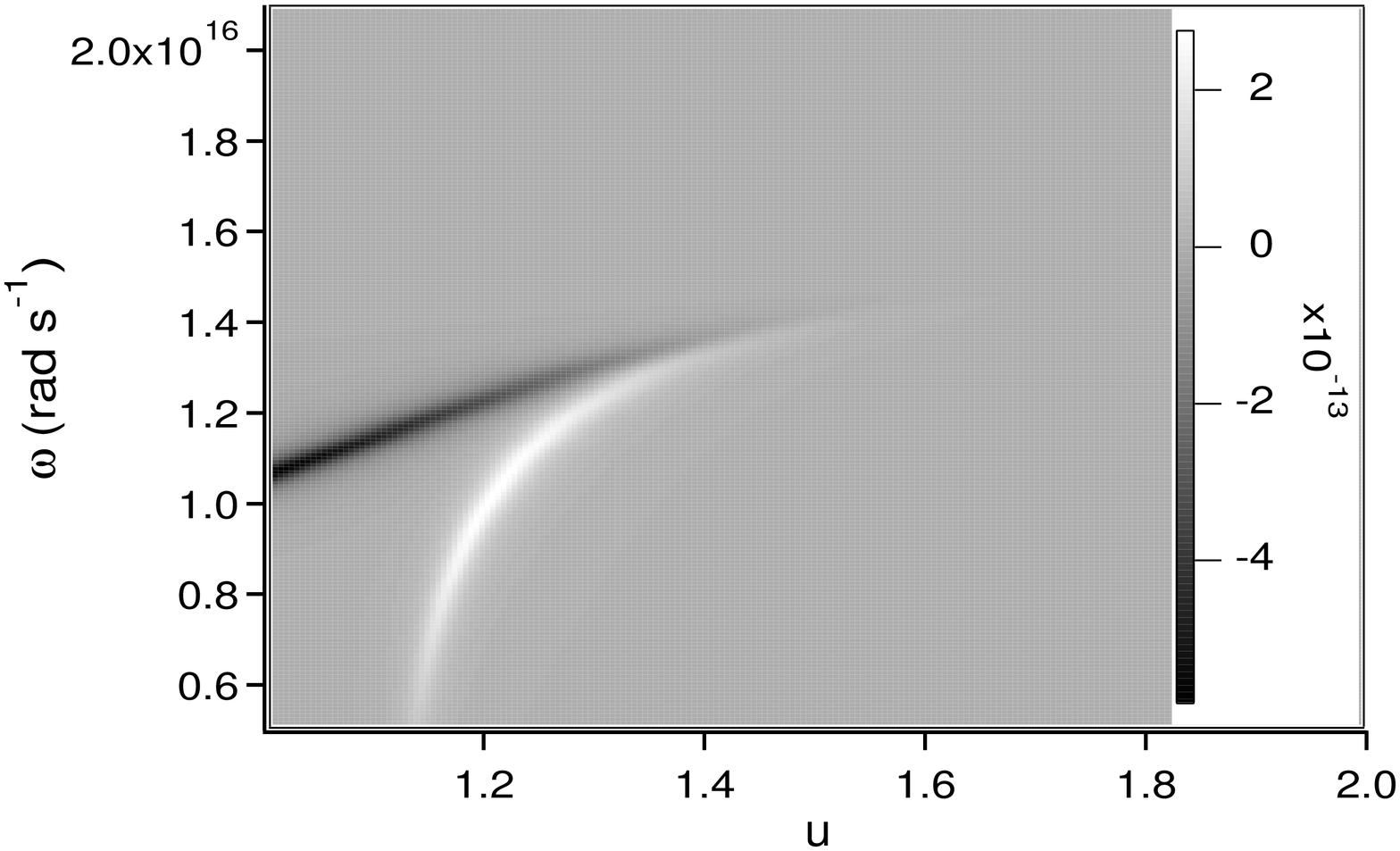}
\includegraphics*[width=\columnwidth,angle=0]{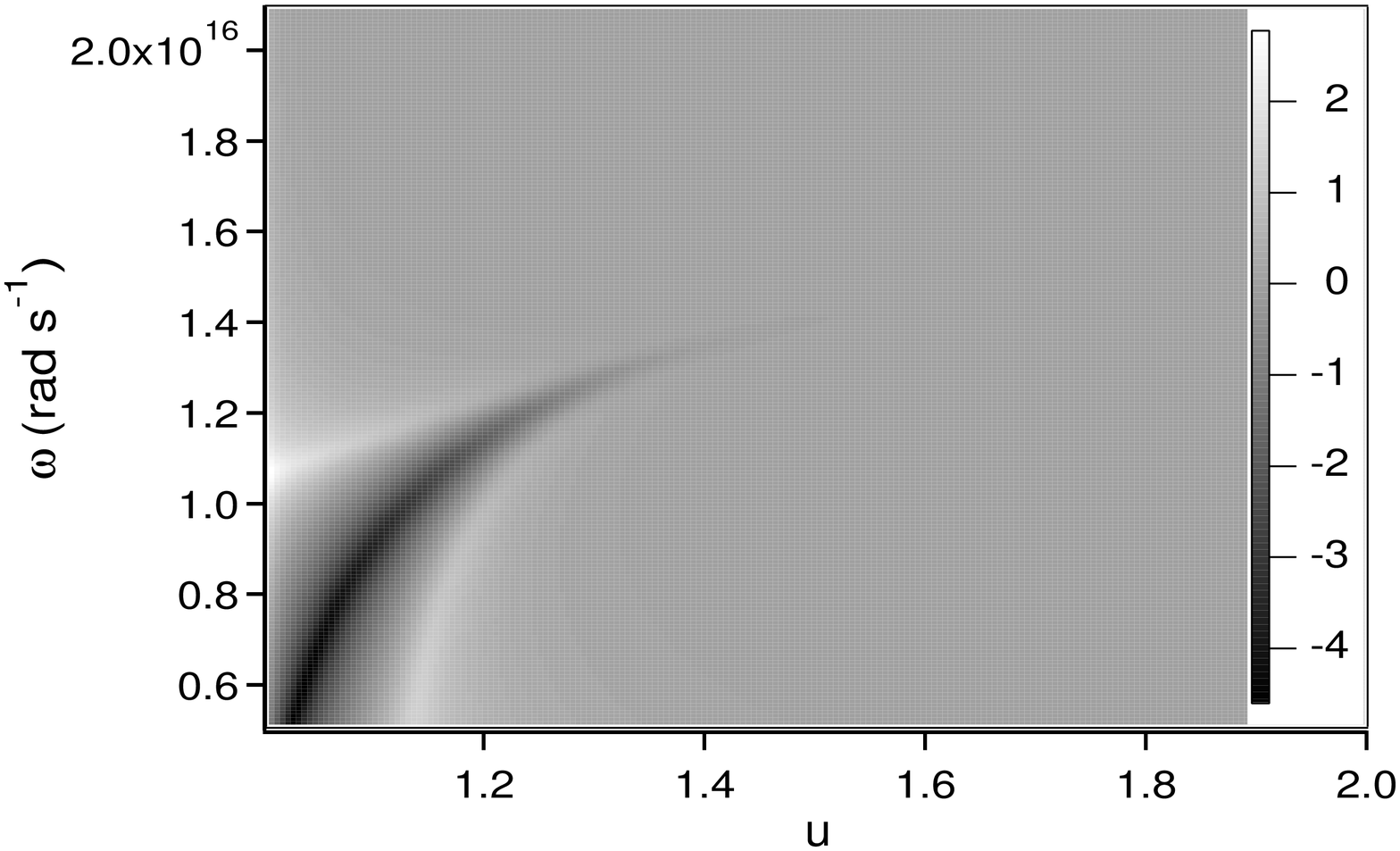}
     \caption{Same as Fig.\ref{fig:2}, but for a separation
$d = 100\,{\rm nm}$.
}
     \label{fig:3}
\end{figure}

In Fig.~\ref{fig:3}, we plot the force for a spacing $d=100$~nm: 
the two branches tend to merge with the flat interface
dispersion relation. 
The following interpretation thus emerges:
when the surfaces approach each other,
the overlapping of the two SPP leads to a splitting of the polariton
frequencies~\cite{Marcuse,GarciaVidal01}.  The frequency splitting can
be found from the solutions of Eq.\,(\ref{eq:spp-pole}) which
are implicitly defined by (see also~\cite{Genet03b})
\begin{equation}
r_{\rm p}(u,\omega) = \pm {\rm e}^{ \omega  v d / c}.
\label{eq:3}
\end{equation}
The signs correspond to either symmetric or antisymmetric mode
functions (for the magnetic field), as shown in Appendix~A and
sketched in Fig.~\ref{fig:2}b. 
The symmetric (antisymmetric)
branch corresponds to a lower (higher) resonance frequency,
respectively, similar to molecular orbitals and tunneling doublets
\cite{Messiah}.
%
%
These branches contribute with opposite signs to the Casimir force,
due to the identity
\begin{eqnarray}
     &&\frac{ 2 \,r_{\rm p}^2( \omega, u ) \, {\rm e}^{ -2
     \omega v d } }{
     1 - r_{\rm p}^2( \omega, u ) \, {\rm e}^{ -2
     \omega v d } }
     =
     \nonumber\\
     &&
     \frac{ r_{\rm p}( \omega, u ) \, {\rm e}^{ -
     \omega v d } }{
     1 - r_{\rm p}( \omega, u ) \, {\rm e}^{ -
     \omega v d } }
-
     \frac{ r_{\rm p}( \omega, u ) \, {\rm e}^{ -
     \omega v d } }{
     1 + r_{\rm p}( \omega, u ) \, {\rm e}^{ -
     \omega v d } }
     ,
\label{eq:4}
\end{eqnarray}
where the first (second) term is peaked at the symmetric
(antisymmetric) cavity mode. The symmetry of the resonance mode
function hence determines the attractive or repulsive character of its
contribution to the Casimir force. We show in Appendix~A by evaluating
explicitly the Maxwell stress tensor, that \emph{symmetric modes are
binding} as in molecular physics. 
%

We note that the splitting in Eq.\,(\ref{eq:4}) of the force spectrum
gives meaningful results also after integration because for evanescent
waves, both terms converge separately. We also point out that for a
complex permittivity $\varepsilon( \omega )$ (as required by the
Kramers-Kronig relations for a dispersive material), the SPP
dispersion relation necessarily moves into the complex plane and
is never satisfied in the real $(u, \omega)$-plane, thus excluding any 
singularities of the integral~(\ref{eq:1a}).

\section{Short-distance limit}

The short-distance behaviour of the Casimir force between non-perfect
metals has been computed in~\cite{Klimchitskaya00,Lambrecht00} 
using tabulated data for the dielectric function and integrating
over imaginary frequencies. We show here that these results can also
be recovered with a real frequency calculation. In particular, we prove that
the interaction between SPPs across the vacuum
gap quantitatively accounts for the short-distance Casimir force
derived in~\cite{Lambrecht00}, thus completing qualitative discussions
put forward by Gerlach~\cite{Gerlach71} and Genet, Lambrecht, and 
Reynaud~\cite{Genet03b}.

For definiteness, 
let us adopt a Lorentz-Drude model for the dielectric function
\begin{equation}
     \varepsilon(\omega) = 1 + \frac{ 2(\Omega^2 - \omega_{0}^2) }{
     \omega_{0}^2 - {\rm i} \gamma \omega - \omega^2 }
,
\label{eq:Drude}
\end{equation}
with resonance frequency $\omega_{0}$ and damping coefficient $\gamma$. 
The corresponding plasma frequency is 
$[2(\Omega^2 - \omega_{0}^2]^{1/2}$.
With this convention, the large $u$ asymptote of
the SPP dispersion~(\ref{eq:2}) occurs at $\omega \approx \Omega$. 
This model can be used to describe either dielectrics or metals when 
$\omega_0=0$.
In the region of large wavevectors, the p-polarized reflection
coefficient has a pole at $\Omega$:
\begin{equation}
     u \gg 1: \quad
r_{\rm p}( \omega, u ) \approx
\frac{ \varepsilon( \omega ) - 1 }{ \varepsilon( \omega ) + 1 }
=
\frac{ \Omega^2 - \omega_{0}^2 }{ \Omega^2 - {\rm i} \gamma \omega - \omega^2 }
.
\label{eq:rp-estat}
\end{equation}
From Figs.\ref{fig:1a}, \ref{fig:1b}, 
we know that the force is significant only in a range around the SPP
resonance. It follows that
the model for $\varepsilon( \omega )$ is needed only in this limited range. 
We have checked that Eq.\,\ref{eq:rp-estat} with $\omega_0=0$ 
is well suited to describe
the reflection data computed from tabulated data for aluminum. 
Note that the results of the fitted parameters ($\Omega$ and $\gamma$
are indicated in the caption of fig.\ref{fig:2})
differ from the usual bulk plasma frequency and damping rates 
that we would get from a fit over the entire spectrum.

\begin{figure}
\includegraphics[width=8cm,angle=0]{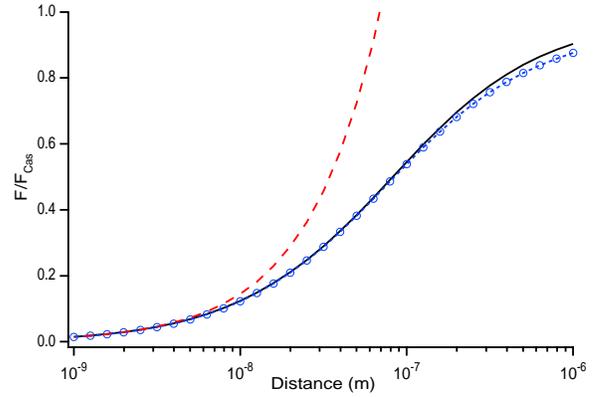}
     \caption{%
Comparison of different expressions for the Casimir 
force between aluminum surfaces. We plot the ratio $F(d) / F_{\rm Cas}(d)$
where $F_{\rm Cas}( d ) = \hbar c \pi^2 / (240 d^4)$ is the
Casimir force for perfect mirrors.
Solid line: numerical integration of Eq.\,(\ref{eq:1}), using
tabulated optical data~\cite{Palik,numerics}.
Short-dashed line with circles: 
same with a model dielectric function of Drude form
[Eq.\,(\ref{eq:Drude})] with $\omega_0 = 0$, 
$\Omega = 1.66 \times 10^{16}\,{\rm s}^{-1}$, and
$\gamma / \Omega = 0.036$. These parameters have been obtained
from a plot of the reflection coefficient 
$(\varepsilon(\omega) - 1)/(\varepsilon(\omega) + 1)$ based on
the tabulated data that has been fitted to the form given in 
Eq.\,(\ref{eq:rp-estat}).
Long-dashed line: short-distance 
asymptotics~(\ref{eq:7}) with the same values for 
$\omega_0, \, \Omega, \, \gamma$.}
\label{fig:force}
\end{figure}

We have checked that 
this formula is well suited to describe the reflection coefficient
computed from tabulated optical data in the frequency region around the
SPP resonance. For aluminum, we get a good agreement
with the values given in the caption of Fig.\ref{fig:force}. These
values do not correspond, of course, to the usual bulk plasma frequency
and damping rates that enter in the Drude model of the dielectric function 
at low frequencies.

With this form of the reflection coefficient, Eq.\,(\ref{eq:3}) yields 
the following dispersion relation
for the (anti)symmetric SPP resonances, neglecting for the
moment the damping coefficient~$\gamma$:
\begin{equation}
     \omega_{\pm}^2 \approx \Omega^2 \left( 1 \mp
     {\rm e}^{ - \omega_{\pm} u d / c } \right)
.
\end{equation}
We have used $v \approx u$ for $u \gg 1$. For large $u$, we solve by
iteration and find that $\omega_{\pm}\,
\raisebox{0.65ex}{$<$}\kern-1.7ex\raisebox{-0.65ex}{$>$}\,
\Omega$. As announced above, the symmetric mode (upper sign) occurs at a
lower resonance frequency.

To derive an analytical estimate for the Casimir force, we retain in
Eq.\,(\ref{eq:1}) only the contribution of p-polarized, evanescent
waves, containing the SPP resonance. 
Introducing the new variable $x =
\omega v d / c$, we get using the identity~(\ref{eq:4})
\begin{eqnarray}
     F &=&
     \frac{\hbar}{4\pi^2 d^3} \;
     {\rm Im}
     \int_{0}^{\infty}\!{\rm d}\omega
         \int_{0}^{\infty}\! x^2{\rm d}x 
	\times
	\nonumber\\
     &&
     \sum_{\lambda \,=\,\pm 1}
     \frac{ \lambda \,{\rm e}^{-x} }{
     r_{\rm p}^{-1}( \omega, c x/(\omega d) ) - \lambda \, {\rm e}^{ -x } }
     ,
\end{eqnarray}
where $\lambda = \pm 1$ corresponds to symmetric (antisymmetric)
modes, respectively.  The integral is dominated by the range $x \sim
1$ and $\omega \sim \Omega$. To leading order in $\Omega d / c \to 0$, we
can thus use the asymptotic form of $r_{\rm p}$ valid for large $u$
given by Eq.\,(\ref{eq:rp-estat}). Performing the integral over
$\omega$ analytically and including damping to first order in
$\gamma/\Omega$ yields
\begin{eqnarray}
     F &=& \frac{ \hbar \Omega }{ 4\pi \, d^3 }
     \int_{0}^{\infty}\! {\rm d}x \,x^2
     \times
     \nonumber\\
     &&
     \sum_{\lambda \,=\,\pm 1}
     \left(
     \frac{ \lambda z \,{\rm e}^{-x} 
          }{ 2 \sqrt{ 1 - \lambda z \, {\rm e}^{ - x} } }
     -
     \frac{ \gamma \lambda z \,{\rm e}^{-x} 
          }{ 2\pi\Omega (1 - \lambda z \, {\rm e}^{ -x}) }
     \right).
     \label{eq:6}
\end{eqnarray}
where $z = 1 - \omega_0^2 / \Omega^2$.
This result shows clearly that symmetric and antisymmetric modes give
Casimir
forces of opposite sign.  The first term in the parenthesis can be
computed by expanding the square root in a power series in $\lambda
z \,{\rm e}^{ - x}$, leading to an infinite series given
in~\cite{Lambrecht00,Genet03b}. The second term, the correction due to
damping, can be integrated in terms of the polylogarithmic
function, so that we finally have
\begin{equation}
     F = \frac{ \hbar \Omega }{ 4\pi d^3 }
\left( \alpha( z ) -
     \frac{ \gamma \,{\rm Li}_3(z^2) }{ 4\pi\Omega }
     \right),
     \label{eq:7}
\end{equation}
where
\begin{equation}
     \alpha( z ) = \frac14 \sum_{n=1}^{\infty}
     z^{2n}
     \frac{(4n-3)!!}{n^3 (4n-2)!!} 
\end{equation}
and
\begin{equation}
     {\rm Li}_3(z^2) = \sum_{n=1}^{\infty} \frac{ z^{2n} 
}{ n^3 }
.
\end{equation}
For completeness, we give the asymptotic series for small
$\omega_0 / \Omega$ ($z \to 1$)
\begin{eqnarray}
\alpha(z ) & \approx &
0.1388 - 0.32 \,( 1 - z ) + 0.4 \,( 1 - z)^2
\label{eq:alpha-series}
\\
{\rm Li}_3( z^2 ) & \approx &
\zeta(3) - \frac{ \pi^2 }{ 3 } ( 1 - z )
\\
&& {} + 
\left[ 3 - \frac{ \pi^2 }{ 6 } 
- 2 \log( 2 (1-z)) \right] (1 - z)^2
\nonumber
\end{eqnarray}
with $\zeta(3) \approx 1.202$. (The coefficient of the second order
term in Eq.\,(\ref{eq:alpha-series}) is only accurate up to
a logarithmic correction.)

Our result Eq.\,(\ref{eq:7}) for the short-distance Casimir force 
agrees with the formula given in~\cite{Lambrecht00,Genet03b} in the
special case $\gamma = 0$, $\omega_0 = 0$ (lossless Drude model).
A very similar expression has been found in~\cite{Raabe03a}.
We compare Eq.\,(\ref{eq:7}) in Fig.~\ref{fig:force} to the full integral 
Eq.\,(\ref{eq:1}) for the case of aluminum: 
it turns out to be
quite accurate for distances $d \le 0.1 \, \lambda_{\rm SPP}$ where
$\lambda_{\rm SPP} = 115$~nm is the wavelength of the SPP
with the largest frequency
\cite{numerics}. In the case of aluminum, the first order correction
in $\gamma/\Omega$ is 2.5\% of the zeroth order value of the force.
The plot also shows that for the numerical integration, the tabulated data
and the Lorentz-Drude model~(\ref{eq:Drude}) with parameters fitted
around the surface resonance give very close results over a large range of
distances. 
This is another indication that the short-range Casimir force 
between real metals is dominated by a narrow frequency range.
Differences of the order of a few percent appear at large
distances where the Casimir force is dominated by the low-frequency
behaviour of the reflection coefficient that is not accurately
modelled with the fitted parameters. 


%

We finally note that the correction of order $\gamma / \Omega$ derived
here 
introduces the effects of losses and must not be confused with the 
correction due to a finite real permittivity
This is already taken into account by the finite value 
of the plasma frequency
$\Omega$ and is responsible for the emergence of the short-distance
regime $d \ll \lambda_{\rm SPP}$
where the Casimir force $\sim 1/d^3$~\cite{Lifshitz56}. At
large distances, a finite $\Omega$ leads to a small correction to the
well-known Casimir force $\sim 1/d^4$ between perfect
conductors~\cite{Schwinger78,Klimchitskaya00,Lambrecht00}.

\section{Conclusion}

We have pointed out that the Casimir attraction between realistic
materials can be quantitatively understood, at short distances, in
terms of the interaction between electromagnetic surface plasmon (or
phonon) polaritons. The modes overlap across the vacuum gap and split into
symmetric and antisymmetric combinations which contribute with
different signs to the Maxwell stress tensor and hence to the Casimir
force. We discussed in particular the short-distance regime
of the Casimir force where $F = H / d^3$ and have given an analytical formula
for the Hamaker constant $H$. We recover previous results
for nonabsorbing materials and evaluate a correction due to
absorption.  Our results have been validated by comparing to a
numerical calculation based on Lifshitz theory.

The approach presented here has the advantage of linking
in a transparent way the Casimir force to the actual
physical properties of the material surface. This suggests
the possibility of engineering the surface plasmon polariton
dispersion relation to modify the Casimir force. Indeed, as it has been shown,
the Casimir force at short distances is entirely due to the interaction between
surface polaritons. Magnetic materials which exhibit Casimir repulsion
\cite{repulsive}
and support s-polarized surface waves when Re $\mu < -1$ \cite{Ruppin00}
are good candidates. The folding of the dispersion relation in reciprocal
space by a grating, known to change the surface wave behaviour
\cite{greffet} could also lead to a substantial modification of the
Casimir force.



\smallskip

\noindent{\it Acknowledgments.}
--- This work has been supported by the bilateral French-German
programme ``Procope'' under project numbers 03199RH and D/0031079.

\appendix

\section{Angular spectrum analysis}

In this appendix, we compute the Casimir force in terms of an
angular spectrum representation of the electromagnetic fields
that is adapted to the planar geometry at hand.

Letting the vacuum gap occupy the region $-d < z < 0$, we can
expand the electric field in the gap as
\begin{eqnarray}
{\bf E}( {\bf x}, \omega ) &=&
\sum_{\mu \,=\, {\rm s},\,{\rm p}}
\int\!d^2K \,\left(
E_-^\mu( {\bf K} ) \,{\rm e}^{ - i k_z z }{\bf e}_\mu^-
+
\right.
\nonumber\\
&& \qquad\left.
E_+^\mu( {\bf K} ) \,{\rm e}^{ i k_z (z+d) }{\bf e}_\mu^+
\right)
\,{\rm e}^{i {\bf K}\cdot{\bf X} }
\label{eq:E-ang-spec}
\end{eqnarray}
where ${\bf K} = (k_x,k_y)$ is the component of the wavevector
parallel to the interfaces and $k_z = \sqrt{ (\omega/c)^2 - K^2 }$ its
perpendicular component. The ${\bf e}_\mu^\pm$ ($\mu = $ s, p) are 
unit polarization vectors, and $E_\pm^\mu( {\bf K} )$ are the
amplitudes of up- and downwards propagating plane waves. A similar
expansion holds for the magnetic field ${\bf H}( {\bf x}, \omega )$
with amplitudes $H_\pm^\mu( {\bf K} )$. We get the averaged Maxwell
stress tensor by integrating incoherently over the contributions
$T_{zz}^{\mu}( {\bf K} )$ of individual modes. For the particular case
of a p-polarized evanescent mode ($K > \omega$), we get by
straightforward algebra
\begin{equation}
T_{zz}^{\rm p}( {\bf K} ) = 2 \mu_0 v^2
\,
{\rm Re}\left[ H_+^{{\rm p}*}( {\bf K} ) H_-^{\rm p}( {\bf K} )
\right]
.
\label{eq:Tzz-per-mode}
\end{equation}
The up- and downward propagating amplitudes are of course related
via the reflection coefficient from the upper interface. Taking
the phase references in Eq.\,(\ref{eq:E-ang-spec}) into account,
we have
\begin{equation}
H_-^{\rm p} = r_{\rm p} \,{\rm e}^{ i k_z d }  H_+^{\rm p}
= r_{\rm p} \,{\rm e}^{ - \omega  v d / c } H_+^{\rm p} 
\approx \pm H_+^{\rm p},
\end{equation}
where the last equality applies in the vicinity of the coupled surface
resonances defined by Eq.\,(\ref{eq:3}). The condition $r_{\rm p} \,{\rm e}^{ -
\omega v d / c } = +1$ thus corresponds to a symmetric magnetic
field distribution on both interfaces, because $H_+^{\rm p}=H_-^{\rm
p}$. In addition, with our sign convention, this mode gives an
attractive contribution proportional to $+|H_-^{\rm p}|^2$ to the
stress tensor~(\ref{eq:Tzz-per-mode}). The opposite is true for
antisymmetric modes.



The sign of the Casimir force due to the coupled polariton
modes can also be understood in terms of the charge densities 
excited on the surfaces, as pointed out by Gerlach~\cite{Gerlach71}. 
These can be found from the normal component of the electric field. 
For a symmetric mode, we get surface charges with opposite sign,
hence an attractive force,
while an antisymmetric mode corresponds to equal surface charges.



\begin{thebibliography}{99}

\bibitem{Casimir}
H. B. G. Casimir, Proc.\ Koninkl.\ Ned.\ Akad.\ Wetenschap.\ {\bf 51},
793 (1948)

\bibitem{Schwinger78}
J. Schwinger, J. Lester L.~DeRaad and K.~A.\ Milton,
{Ann.\ Phys.\ (N.Y.)} {\bf 115}, 1 (1978).

\bibitem{Plunien}
G. Plunien, B. M\"uller and W. Greiner, Phys.\ Rep.\ {\bf 134}, 87
(1986) 

\bibitem{Bordag}
M. Bordag, U. Mohideen and V. M. Mostepanenko, Phys.\ Rep.\ {\bf 353},
1
(2001)

\bibitem{Lamoreaux99}
S. K. Lamoreaux, Am.\ J.\ Phys.\ {\bf 67}, 850 (1999)

\bibitem{Lamoreaux97}
S. K. Lamoreaux, Phys.\ Rev.\ Lett.\ {\bf 78}, 5 (1997)

\bibitem{Mohideen}
U. Mohideen and A. Roy, Phys.\ Rev.\ Lett.\ {\bf 81}, 4549 (1998)

\bibitem{Klim}
G. L. Klimchitskaya, A. Roy, U. Mohideen, and V.M. Mostepanenko,
{Phys.\ Rev.\ A} {\bf 60}, 3487 (1999)

\bibitem{Klimchitskaya00}
G. L. Klimchitskaya, U. Mohideen, and V. M. Mostepanenko,
{Phys.\ Rev.\ A} {\bf 61}, 062107 (2000)

\bibitem{Lambrecht00}
A. Lambrecht and S. Reynaud, Eur.\ Phys.\ J.\ D {\bf 8}, 309 (2000)

\bibitem{Genet00}
C. Genet, A. Lambrecht and S. Reynaud,
{Phys.\ Rev.\ A} {\bf 62}, 012110 (2000)


\bibitem{Tadmor01}
R. Tadmor,
J. Phys.: Condens.\ Matt.\ {\bf 13}, L195 (2001)

\bibitem{Genet03a}
C. Genet, A. Lambrecht, P. Maia Neto and S. Reynaud,
Europhys.\ Lett.\ {\bf 62}, 484 (2003)

\bibitem{Chan1}
H. B. Chan, V. A. Aksyuk, R. N. Kleiman, D. J. Bishop and F. Capasso, Phys.\
Rev.\ Lett.\ {\bf 87}, 211801 (2001)

\bibitem{Chan2}
H. B. Chan, V. A. Aksyuk, R. N. Kleiman, D. J. Bishop and F. Capasso,
Science {\bf 291}, 1941 (2001)

\bibitem{Buks}
E. Buks and M. L. Roukes, Phys.\ Rev.\ B {\bf 63}, 033402 (2001)

\bibitem{Milloni}
P. W. Milloni, \emph{The Quantum Vacuum: An Introduction to Quantum
Electrodynamics} (Academic Press, London, 1994)

\bibitem{Mostepanenko}
V.~M.\ Mostepanenko and N.~N.\ Trunov,
\emph{The Casimir Effect and Its Applications}
(Oxford Science Publications, Oxford, 1997)

\bibitem{Lifshitz56}
E. M. Lifshitz,
Soviet Phys.\ JETP {\bf 2}, 73 (1956)
[{J.\ Exper.\ Theoret.\ Phys.\ USSR} {\bf 29}, 94 (1955)]

\bibitem{VanKampen68}
N. G. Van Kampen, B. R. A. Nijboer and K. Schram,
Phys.\ Lett.\ A {\bf 26}, 307 (1968)

\bibitem{Gerlach71}
E. Gerlach, Phys.\ Rev.\ B {\bf 4}, 393 (1971)

\bibitem{Genet03b}
C. Genet, A. Lambrecht and S. Reynaud,
preprint quant-ph/0302072 (2003)

\bibitem{Palik}
E.D. Palik, \emph{Handbook of Optical constants of Solids},
(Academic Press, San Diego, 1991)

\bibitem{Rytov3}
S. M. Rytov, Yu. A. Kravtsov and V. I. Tatarskii,
\emph{Elements of Random Fields}, vol.\ 3 of
\emph{Principles of Statistical Radiophysics}
(Springer, Berlin, 1989)

\bibitem{Scheel00b}
L. Kn{\"o}ll, S. Scheel and D.-G. Welsch,
\emph{QED in Dispersing and Absorbing Media}, in
\emph{Coherence and Statistics of Photons and Atoms},
edited by J. Perina (John Wiley \& Sons, Inc., New York, 2001)

\bibitem{Raabe03a}
C. Raabe, L. Kn\"oll, and D.-G. Welsch,
Phys. Rev. A 68 (2003) 033810. Note that Eq.\,(82) in this paper
is based on an inaccurate
evaluation of the integral Eq.\,(E9). Once this is corrected, we find 
agreement with Eq.\,(\ref{eq:7}) reported here.

\bibitem{Henkel02a}
C. Henkel, K. Joulain, J.-Ph. Mulet and J.-J. Greffet,
J.\ Opt.\ A: Pure Appl.\ Opt.\ {\bf 4}, S109 (2002)


\bibitem{Ford1}
L. H. Ford, Phys.\ Rev.\ D {\bf 38}, 528 (1988)

\bibitem{Ford2}
L. H. Ford, Phys.\ Rev.\ A {\bf 48}, 2962 (1993)

\bibitem{Raether}
H. Raether,
\emph{Surface Plasmons on Smooth and Rough Surfaces and on Gratings}
(Springer, Berlin, 1988)

\bibitem{Shchegrov}
A.V. Shchegrov, K. Joulain, R. Carminati and J.-J. Gref\-fet,
Phys. Rev. Lett, {\bf 85}, 1548 (2000)

\bibitem{Joulain03}
K. Joulain, R. Carminati, J.-Ph. Mulet, and J.-J. Greffet,
Phys.\ Rev.\ B (2003), in press; e-print physics/0307018.

\bibitem{Marcuse}
D. Marcuse,
\emph{Theory of Dielectric Optical Waveguides},
2nd ed.\ (Academic Press, San Diego, 1991)

\bibitem{GarciaVidal01}
A. Krishnan, T. Thio, T. J. Kim, H. J. Lezec, T. W. Ebbesen,
P. A. Wolff, J. Pendry, L. Martin-Moreno and F. J. Garcia-Vidal,
Opt.\ Commun.\ {\bf 200}, 1 (2001)

\bibitem{Messiah}
A. Messiah, \emph{M\'ecanique quantique}, vol.\ 1, new
   ed.\ (Dunod, Paris, 1995).

\bibitem{numerics}
The numerical integration uses the Lifshitz formula and proceeds
along the imaginary frequency axis $\omega = {\rm i}\xi$. 
The dielectric function $\varepsilon( {\rm i}\xi)$ is constructed
from the tabulated data at real frequencies using the sum rules
given in~\cite{Lambrecht00,Klimchitskaya00}.
%

\bibitem{repulsive}
O. Kenneth, I. Klich, A. Mann and M. Revzen,
Phys.\ Rev.\ Lett.\ {\bf 89}, 033001 (2002)

\bibitem{Ruppin00}
R. Ruppin,
Phys.\ Lett.\ A {\bf 277}, 61 (2000)

\bibitem{greffet}
J.-J. Greffet, R. Carminati, K. Joulain, J.-Ph. Mulet, S. Mainguy, 
and Y. Chen,
Nature, {\bf 416}, 61 (2002) 

\end{thebibliography}
\end{document}